\documentclass[11pt,a4paper]{article}

\usepackage{amsthm}
\usepackage{amssymb,amsmath,latexsym}
\usepackage{amsfonts}
\usepackage{epsfig}
\usepackage{graphicx}
\usepackage{tikz} 

\usepackage{mathdots}

\usepackage{a4}

\newtheorem{theorem}[equation]{Theorem}

\newtheorem{example}[equation]{Example}

\theoremstyle{definition}
\newtheorem{definition}[equation]{Definition}
\newtheorem{remark}[equation]{Remark}
\newtheorem{problem}[equation]{Problem}

\numberwithin{equation}{section}

\newcommand{\R}{{\mathbb{R}}}

\newcommand{\N}{{\mathbb{N}}}

\def\nref#1{{\rm (\ref{#1})}}

\begin{document}

\title{Controlling centrality: The Inverse ranking problem for spectral centralities of complex networks}

\author{Esther Garc\'{\i}a$^1$ and Miguel Romance$^{1,2,3}$
\\
{\small $^1$ Departamento de Matem\'atica Aplicada,}\\
{\small Ciencia e Ingenier\'{\i}a de los Materiales y Tecnolog\'{\i}a Elctr\'onica,}\\
{\small  ESCET - Universidad Rey Juan Carlos, 28933 M\'ostoles (Madrid), Spain}
\\
{\small $^2$ Center for Computational Simulation,}\\
{\small  28223 Pozuelo de Alarc\'on (Madrid), Spain}
\\
{\small $^3$ Laboratory of Math. Computation on Complex Networks and their Applications, }\\
{\small Universidad Rey Juan Carlos, 28933 M\'ostoles (Madrid), Spain}
}

\maketitle

\begin{abstract}
In this paper some results about the controllability of spectral centrality in a complex network are presented. In particular, the inverse problem of designing an unweigthed graph with a prescribed centrality is considered, by showing that for every possible ranking eventually with ties, an unweighted directed/undirected complex network can be found whose PageRank or eigenvector centrality gives the ranking considered. Different families of networks are presented in order to analytically solve this problem either for directed and undirected graphs with and without loops. 
\end{abstract}


\vspace{1cm}
\hrule

\section{Introduction}\label{sec:intro}

Some of the main goals of Science are understanding, predicting and controlling systems coming from the real life \cite{Cohen,Eisner}. If we consider a real phenomenon, the first step to be considered is understanding how it works, what are its basic ingredients, the causes and effects and the mechanisms  behind it. In a second level of knowledge, a proper scientific modelization allows us to predict the results and effects of the real phenomenon in advance. Finally, in the last frontier of knowledge, we should be able to efficiently modify the contour variables of the phenomenon in order to obtain and control the results derived from it \cite{Schraw}. From a mathematical point of view, there are many problems that have been mastered enough in order to control their results at our wish, including the evolution of dynamical systems modeled with ordinary and partial differential equations on general systems \cite{Glasser,ST2000}, but in this paper we focus on controlling the structural properties of complex networks.

Complex networks have widely shown to be a sound mathematical model that fits many real applications, from biological to technological systems, passing through many other real applications, such as social systems, economics, linguistics and many other \cite{B2006, Newman, Estrada}. During the last decades the scientific community of Network Science have casted a huge development of models, tools and multidisciplinary techniques in order to understand, predict and control many real systems that can be understood by using complex networks \cite{B2006, Newman, Estrada}. If we focus on controlling systems modeled by some kind of complex networks, most part of the efforts done by the scientific community so far have been concentrated on mastering dynamical processes that take place on complex networks \cite{Meyn, RR2014, LSB2011} by extending the classic results from Control theory  \cite{Glasser,ST2000}, but much less attention have been attached to other control problem dealing with different problems in complex networks analysis. One of these control problems includes the control of some structural parameters of a complex network, such as centrality measures \cite{NCRRL12}.

Among the vast set of structural measures of complex networks considered by the scientific community, centrality measures play a crucial roles since they quantify the relevance of the nodes in the structure and dynamics on complex networks \cite{B2006,Lu2016}. There is no consensus about a single centrality measure that resumes the relevance of each node and therefore there are many complementary measures that quantify different aspects of the role of each node, such as the degree, the eigenvector centrality, PageRank, HITS-Authority score, betweenness centrality, among many others \cite{B2006,Lu2016,Newman, Estrada}.

It is natural considering the control problem related to centrality measures, i.e. given complex network $\mathcal{G}=(\mathcal{N},\mathcal{E})$ and a centrality measure $c:\mathcal{N}\longrightarrow \R$, we want to modify $\mathcal{G}$ in order to get its centrality measure $c(\mathcal{\tilde G})$ at our wish. This problem was considered for the eigenvector centrality \cite{Bonacich,BLl2001} of weighted networks by V.\,Nicosia et al. \cite{NCRRL12}, by proving that given a weighted and strongly connected complex network $\mathcal{G}=(\mathcal{N},\mathcal{E})$ with $n\in\N$ nodes, then for every vector $\mathbf{v}=(v_1,\cdots,v_n)^t\in\R^n$  with $v_i\ge 0$ for all $1\le i\le n$ and $v_1+\cdots+v_n=1$ we can modify the weights in $\mathcal{G}$ in order to get that its eigenvector centrality  is $\mathbf{v}$. In particular, given a vector $\mathbf{v}=(v_1,\cdots,v_n)^t\in\R^n$  with $v_i\ge 0$ for all $1\le i\le n$ and $v_1+\cdots+v_n=1$ it is  easy to spot a weighted (and strongly connected) network $\mathcal{G}=(\mathcal{N},\mathcal{E})$ such that its eigenvector centrality or its PageRank centrality\cite{BP1998,PaBrMoWi} is $\mathbf{v}$. The main goal of this paper is considering this control problem for the eigenvector \cite{Bonacich,BLl2001} and PageRank \cite{BP1998,PaBrMoWi}, but for unweighted networks. Note that if we restrict ourselves to unweighted networks and we only allow modifying  the structure by adding/removing links, then by using an easy cardinality argument, for any vector  in general $\mathbf{v}=(v_1,\cdots,v_n)^t\in\R^n$  with $v_i\ge 0$ for all $1\le i\le n$ and $v_1+\cdots+v_n=1$ there is no unweighted network $\mathcal{G}=(\mathcal{N},\mathcal{E})$  whose eigenvector or PageRank centrality is $\mathbf{v}$, since there is a finite number of unweighted networks with $n$ nodes but an infinite number of vectors $\mathbf{v}=(v_1,\cdots,v_n)^t\in\R^n$  such that $v_i\ge 0$ for all $1\le i\le n$ and $v_1+\cdots+v_n=1$. 

As an alternative, we consider the following control problem related not with the precise value of the centrality measure, but with the ranking of nodes according to its centrality. Given a ranking (possibly with ties) of the nodes of the network  (i.e. given an ordering $d_{\sigma(1)}\succsim d_{\sigma(2)}\succsim \cdots \succsim d_{\sigma(n)}$, for some permutation $\sigma$ of the node set and $\succsim$ means sometimes $=$ and sometimes is $>$), the goal is spotting an unweighted network such that if we reorder its nodes by using the centrality of each node (descent ordering), then we recover the prefixed  ranking (possibly with ties). This is what we will call later the {\sl inverse ranking problem}. Note that in many applications, the information given by the ranking of centrality measures is even more valuable that the actual value of the centrality measure. An example of this fact can be found, for example in the use of PageRank for web engines, where it is more important to get a higher position in the web ranking than the actual value of the PageRank of each webpage \cite{LM2011}.

By using the results obtained by V.\,Nicosia et al. \cite{NCRRL12} it is straightforward that for any descending ordering (possibly with ties) of the nodes of the network we can find a weighted  complex network such that if we reorder its nodes by using the eigenvector centrality or PageRank centrality of each node (descent ordering), then we recover the prefixed  ordering. In this paper, we will consider this inverse centrality problem, but for unweighted networks and takins the eigenvector centrality or the PageRank as centrality measures. 

The structure of the paper is as follows. After this introduction, Section~\ref{sec:def} is devoted to present the basic definitions about centrality measure, rankings and the precise statement of the control problems considered later.  Then, Section~\ref{sec:und-loop} and Section~\ref{sec:direct} include the solution of the inverse ranking problem for eigenvector and PageRank centralities in the case of undirected and directed unweighted networks respectively.

\section{Notation, basic concepts and main problems considered}\label{sec:def}

Let ${\mathcal G}= ({\mathcal N}, {\mathcal E})$ be a (directed or undirected) unweighted graph where ${\mathcal N} = \{ 1,2, \ldots, n \}$ and $n\in \mathbb{N}$.  If ${\mathcal G}$ is an undirected graph, the links are unordered couples of nodes $\{i,j\}$ (possibly $i=j$), while if ${\mathcal G}$ is a directed graph, the links are  ordered couples of nodes  $(i,j)$ with $i,j\in E$ (possibly $i=j$). The {\it adjacency matrix} of ${\mathcal G}$ is an $n\times n$-matrix real matrix
\[
A=(a_{ij})\in M_{n\times n}(\R),\quad \hbox{ where } a_{ij}=\left\{
       \begin{array}{ll}
         1, & \hbox{if $\{i,j\}$ (resp. $(i,j)$) is a link of ${\mathcal G}$} ,\\
         0, & \hbox{otherwise,}
       \end{array}\right.
\]
where $M_{n\times n}(\R)$ denotes the set of all $n\times n$ real matrices. Notice that the adjacency matrix of an undirected graph is necessarily symmetric, while this is not true in general  for directed graphs. For directed graphs, a link $(i,j)$ is said to be an {\it outlink} for node $i$ and an {\it inlink} for node $j$. We denote $gr_{out}(i)$ the {\it outdegree} of node $i$, i.e.,  the number of outlinks of a node $i$. Notice that $gr_{out}(i)=\sum_k a_{ik}$. If $\mathcal{G}$ is undirected, then the outlinks of a node coincide with its inlinks and the outdegree is called simply the degree on the node, denoted by $gr(i)=gr_{out}(i)$.

Vectors of $\mathbb{R}^n$ will be denoted throughout this paper by column matrices. In particular, we will consider
\[
\begin{split}
\mathbf{e}_1&=\left(
                \begin{array}{c}
                          1 \\
                          0\\
                          \vdots \\
                          0 \\
                \end{array}
            \right)=(1,0,\cdots, 0)^t\in \mathbb{R}^n, \ \dots,\
\mathbf{e}_n=\left(
                \begin{array}{c}
                          0 \\
                          0\\
                          \vdots \\
                          1 \\
                \end{array}
            \right)=(0,0,\cdots, 1)^t\in \mathbb{R}^n,\quad\\
\mathbf{e}&=\mathbf{e}_1+\dots+\mathbf{e}_n=\left(
                \begin{array}{c}
                          1 \\
                          1 \\
                          \vdots \\
                          1 \\
                \end{array}
            \right)=(1,1,\cdots, 1)^t,
\end{split}
\]
where $\mathbf{v}^t$ denotes the transposed of vector $\mathbf{v}\in\R^n$. The $i^{\rm th}$-component of a vector $\mathbf{v}=(v_1,\cdots,v_n)^t\in\R^n$ is given by the usual scalar product $\langle \mathbf{v},\mathbf{e}_i\rangle= \mathbf{v}^t\mathbf{e}_i=v_i$, and the sum of the components of the vector $\mathbf{v}\in \mathbb{R}^n$ can be computed as  $\langle \mathbf{v},\mathbf{e}\rangle=\mathbf{v}^t\mathbf{e}$. Moreover, we will say that a vector $\mathbf{v}>0$ if all the components $v_i$ of $\mathbf{v}$ are greater than zero, i.e., $\mathbf{v^t }\mathbf{e}_i>0$, for all $i=1,\dots, n$.

Since we are going to study the inverse ranking problem for some centrality measures, we must start by recalling the basic definitions of the centrality measures considered. As we pointed out in the Introduction, there is no consensus about a single centrality measure that resumes the relevance of each node and therefore there are many complementary measures that quantify different aspects of the role of each node, such as the degree, the eigenvector centrality, PageRank, HITS-Authority score, betweenness centrality, among many others \cite{B2006,Lu2016,Newman, Estrada}. In this paper we will consider two (spectral) centrality measures that are related with the eigenvectors of some matrices related with the graph (mainly the adjancency matrix an some of its renormalizations): the eigenvector centrality \cite{Bonacich,BLl2001}  and the (classic) PageRank centrality \cite{PaBrMoWi}.

{\it Eigenvector centrality} was introduced by P.\,Bonacich\cite{Bonacich,BLl2001} in social network analysis by considering that an individual's centrality must be a function of the centrality of those to whom he or she is connected, so the definition is related with a positive eigenvector of the transposed of the adjacency matrix as follows:

\begin{definition}[\cite{Bonacich,BLl2001}]
Given a (directed or undirected) strongly connected graph ${\mathcal G}= ({\mathcal N}, {\mathcal E})$ with $n$ nodes whose adjacency matrix is denoted by $A\in M_{n\times n}(\R)$, we say that the vector $\mathbf{c}\in\mathbb{R}^n$ is the {\it eigenvector centrality} of the graph ${\mathcal G}$ if $\mathbf{c}$ is the unique (positive) eigenvector with $\langle \mathbf{c},\mathbf{e}\rangle=1$ (i.e. $\mathbf{c}$ is normalized) which associated to the spectral radius $\rho(A)$ of $A$, i.e.,
\begin{equation}\label{eq:eigen}
A^t\mathbf{c}=\rho(A)\mathbf{c}, \quad \mathbf{c}>0, \quad \mathbf{c}^t\mathbf{e}=1.
\end{equation}
Once the eigenvector centrality $\mathbf{c}$ is given, the {\it eigenvector centrality of a node $i\in\mathcal{N}$} is  $c_i=\langle \mathbf{c},\mathbf{e}_i\rangle$.
\end{definition}

Note that the strongly connectedness of $\mathcal{G}$ is the necessary and sufficient condition in order to guarantee the existence and uniqueness of  a positive and normalized eigenvector associated to the spectral radius of $A^t$, simply by using the classic Perron-Frobenius Theorem \cite{Bonacich,BLl2001,Meyer}.

{\it PageRank centrality} was introduced by S.\,Brin and L.\,Page \cite{BP1998,PaBrMoWi} in the context of web searchers, in order to get an efficient algorithm for ranking webpages according to its relevance and it is now at the core of the celebrated {\sf Google} search engine \cite{LM2011}. The basic idea behind this centrality measure is the fact that if we surf at random in the web, {\it the more often we reach a webpage, the more relevant it is}, so in terms of complex networks analysis, the heuristics that support the PageRank centrality is that if we follow a random walker along the network's structure, the centrality of each node is the frequency of passing through it in the Markov chain given by the random walk. Therefore, the mathematical formalization of this idea is given in the following definition:

\begin{definition}[\cite{BP1998,PaBrMoWi,LM2011}]
Consider a (directed or undirected) graph ${\mathcal G}= ({\mathcal N}, {\mathcal E})$ with $n$ nodes whose adjacency matrix is denoted by $A\in M_{n\times n}(\R)$ such that the outdegree of each node is positive, and a damping factor $q\in (0,1)$. Let $P$ be the row-normalized matrix obtained from $A$ and $B=P^t$, i.e.,
\[
B=(b_{ij})\in M_{n\times n}(\R),\quad \hbox{ where }  b_{ij}=\frac{a_{ji}}{\sum_k a_{jk}},
\]
and define
\[
\Phi=\frac{1-q}{n}\mathbf{ee}^t+qB.
\]
We say that the vector $\mathbf{c}\in\mathbb{R}^n$ is the (classic) {\it PageRank centrality} of the graph ${\mathcal G}$ (with damping factor $q$) if $\mathbf{c}$ is the unique (positive) eigenvector with $\langle \mathbf{c},\mathbf{e}\rangle=1$ (i.e. $\mathbf{c}$ is normalized) which associated to the spectral radius of $\Phi$, i.e.,
\[
\Phi\mathbf{c}=\rho(\Phi)\mathbf{c}, \quad \mathbf{c}>0, \quad \mathbf{c}^t\mathbf{e}=1.
\]
Moreover, since  $\Phi$ is a column-stochastic matrix, its spectral radius  $\rho(\Phi)=1$, so $\bf{c}$ satisfies
\begin{equation}\label{eq:PR}
qB\mathbf{c}=\mathbf{c}-\frac{1-q}{n}\mathbf{e},\quad \mathbf{c}>0, \quad \mathbf{c}^t\mathbf{e}=1.
\end{equation}
Once the (classic) PageRank centrality $\mathbf{c}$ is given, the {\it (classic) PageRank centrality of a node $i\in\mathcal{N}$} is  $c_i=\langle \mathbf{c},\mathbf{e}_i\rangle$.
\end{definition}

Note that the PageRank centrality $\mathbf{c}$ with damping factor $q\in(0,1)$ is the steady state of a Markov chain on $\mathcal{G}$ whose transition matrix is $\Phi$ and since $\phi_{ij}>0$ for all $1\le i,j\le n$, then by using the classic Perron Theorem, this steady state always exists and it is unique \cite{BP1998,PaBrMoWi,LM2011}.

If we take a (directed or undirected) graph ${\mathcal G}= ({\mathcal N}, {\mathcal E})$ such that $gr_{out}(i)=0$ for some $i\in\mathcal{N}$ (in this case we say that $i$ is a {\it dangling node}), then  $\Phi$ could be fixed properly in order to extend the definition of PageRank centrality \cite{LM2011}, but we will not need this extension for the purposes of this paper.

There also other more general definitions of (classic) PageRank centrality, such as the {\it personalized Pagerank centrality}\cite{BMPW98,LM2011} with personalization vector $\mathbf{v}\in\R^n$ such that $\langle \mathbf{v},\mathbf{e}_i\rangle>0$ for all $1\le i\le n$ and $\langle \mathbf{v},\mathbf{e}\rangle=1$ (that will be considered later), which is the steady state of a Markov chain on $\mathcal{G}$ whose transition matrix is 
\begin{equation}\label{eq:personalizedPR}
\Phi_{\mathbf{v}}=(1-q)\mathbf{ve}^t+qB.
\end{equation}

Next ingredient needed in order to consider the inverse ranking problem or some centrality measures is the concept of ranking. Roughly speaking, a {\sl ranking (without ties)} of items is a rank-ordered list of the items \cite{LM2012}, but this concept can be mathematically modeled as follows:

\begin{definition}
Given a finite family of items $S=\{s_1,\cdots,s_n\}$, a {\it ranking} of $S$ (without ties)  is given by a bijective application $\phi:S\longrightarrow\{1,\cdots, n\}$ and the rank-ordered list of the items in $S$ is 
\[
s_{\phi^{-1}(1)}\succ s_{\phi^{-1}(2)}\succ\cdots\succ s_{\phi^{-1}(n)}.
\]
\end{definition}

Similarly, a  {\sl ranking (possibly with ties)} of items is a rank-ordered list of the items such that we allow that two items could be at the same positions (i.e. {\sl they are tied}). This idea can be formalized as follows:

\begin{definition}
Given a finite family of items $S=\{s_1,\cdots,s_n\}$, a {\it ranking (possibly with ties)} of $S$ can be represented by a surjective application $\phi:S\longrightarrow\{1,\cdots, k\}$ for some $1\le k\le n$. Note that the rank-ordered list of the items in $S$ is again $s_{\phi^{-1}(1)}\succ s_{\phi^{-1}(2)}\succ\cdots\succ s_{\phi^{-1}(k)}$, but each $s_{\phi^{-1}(j)}$ now corresponds to a subset of items that are tied.
\end{definition}

Note that it is straightforward to point out that in the previous definition, there are some ties in the ranking if and only if $k<n$.  

Among all ranking (possibly with ties) of a finite family, the following definition presents  a subclass of rankings that {\sl do not transform too much} the lexicographical order given in $S=\{s_1,\cdots,s_n\}$. This kind of ranking will play a relevant role in the rest of the paper.

\begin{definition}
Let $S=\{s_1,\cdots,s_n\}$ be a finite family. Given a list of integers $(r_1,\cdots r_k)$ such than $1\le r_1,\cdots,r_k\le n$  and $r_1+\cdots+r_k=n$, the {\it $(r_1\cdots, r_k)$-ranking (possibly with ties)} is the raking given by the surjective application $\phi:S\longrightarrow\{1,\cdots, k\}$ defined as
\[
\begin{split}
\phi(s_1)&=\phi(s_2)=\cdots =\phi(s_{r_1})=1,\\
\phi(s_{r_1+1})&=\phi(s_{r_1+2})=\cdots =\phi(s_{r_1+r_2})=2,\\
&\cdots\\
\phi(s_{r_1+\cdots+r_{k-1}+1})&=\phi(s_{r_1+\cdots+r_{k-1}+2})=\cdots =\phi(s_{n})=k.\\
\end{split}
\]
\end{definition}

It is easy to check that a $(r_1\cdots, r_k)$-ranking of a family $S=\{s_1,\cdots,s_n\}$ has no ties if and only if $k=n$ and in this case $r_1=\cdots=r_n=1$, so the raking obtained is
\[
s_1\succ s_2\succ \cdots \succ s_n.
\]

Once we have introduced all the notations and basic concepts, we state precisely all the problems that will be considered in the next sections of the paper. As it was pointed out in the Introduction, by using the results obtained by V.\,Nicosia et al.\cite{NCRRL12}, it is easy to give a positive answer to the following problem for the eigenvector centrality in weighted networks:

\begin{problem}[Inverse centrality problem for weighted networks]
Given a centrality measure and vector $\mathbf{v}=(v_1,\cdots,v_n)^t\in\R^n$  with $v_i\ge 0$ for all $1\le i\le n$ and $v_1+\cdots+v_n=1$, find a weighted network $\mathcal{G}=(\mathcal{N},\mathcal{E})$ such that its centrality is $\mathbf{v}$.
\end{problem}

It is straightforward to check that if we restrict ourselves to unweighted networks and we only allow modifying the structure by adding/removing links, then since there is a finite number of unweighted networks with $n$ nodes but an infinite number of vectors $\mathbf{v}=(v_1,\cdots,v_n)^t\in\R^n$  such that $v_i\ge 0$ for all $1\le i\le n$ and $v_1+\cdots+v_n=1$, in general there is no unweighted network $\mathcal{G}=(\mathcal{N},\mathcal{E})$  whose eigenvector or PageRank centrality is $\mathbf{v}$.

As an alternative, we can consider the weaker problem related with rankings and that has sense for unweigthed networks either for eigenvector centrality or PageRank centrality:

\begin{problem}[Inverse ranking problem for unweighted networks]
Given a centrality measure such as eigenvector or PageRank centrality and a ranking (possibly with ties) of the nodes of the network, find an unweighted network $\mathcal{G}=(\mathcal{N},\mathcal{E})$ such that  if we reorder its nodes by using the centrality of each node (descent ordering), then we recover the prefixed  ranking (possibly with ties).
\end{problem}

Since this problem is weaker than the Inverse centrality problem, it is straightforward to point out that the Inverse ranking problem for weighted networks has positive answer for igenvector or PageRank centralities.

Despite the fact that the Inverse ranking problem is weaker than the Inverse centrality problem, note that in many applications, the information given by the ranking of centrality measures is even more valuable that the actual value of the centrality measure. An example of this fact can be found, for example in the use of PageRank for web engines, where it is more important to get a higher position in the web ranking than the actual value of the PageRank of each webpage \cite{LM2011}.

Note that the Inverse ranking problem can be rewritten in terms of $(r_1\cdots, r_k)$-rankings as the following theorem shows.

\begin{theorem}\label{thm:equivalence}
Given a centrality measure, the following assertions are equivalent:
\begin{itemize}
 \item[{\it (i)}] The inverse ranking problem has positive solution for every ranking (possibly with ties).
 \item[{\it (ii)}] For every possible $(r_1\cdots, r_k)$, the inverse ranking problem has positive solution for all $(r_1\cdots, r_k)$-rankings (possibly with ties).
\end{itemize}
\end{theorem}

\begin{proof}
\noindent {\it (i)}$\implies${\it (ii)} Is trivial since {\it (ii)} is weaker than {\it (i)}. For the reverse, note that it we take a ranking (possibly with ties) $r$ of the set of nodes $\mathcal{N}=\{1,\cdots, n\}$, by using its definition there is a a surjective application $\phi:\mathcal{N}\longrightarrow\{1,\cdots, k\}$ for some $1\le k\le n$, such that the ranking is given by
\[
\phi^{-1}(1)\succ \phi^{-1}(2)\succ\cdots\succ \phi^{-1}(k),
\]
where each $\phi^{-1}(j)$ corresponds to a subset of nodes that must be tied. Let us denote $r_j=|\phi^{-1}(j)|$ for every $1\le j\le k$ and take a permutation $\sigma: \mathcal{N}\longrightarrow \mathcal{N}$ such that
\[
\begin{split}
\phi^{-1}(1)&=\sigma(\{1,2,\cdots,r_1\}),\\
\phi^{-1}(2)&=\sigma(\{r_1+1,r_2+2,\cdots,r_1+r_2\}),\\
&\cdots\\
\phi^{-1}(k)&=\sigma(\{r_1+\cdots+r_{k-1}+1,\cdots,n\}).\\
\end{split}
\]
Now, by hypothesis, there is an unweighted network $\mathcal{G}=(\mathcal{N},\mathcal{E})$ such that  if we reorder its nodes by using the centrality of each node we recover the $(r_1\cdots, r_k)$-ranking. Finally, if we take  $\mathcal{\tilde G}$ by permuting the nodes of $\mathcal{G}$ with $\sigma$, then $\mathcal{\tilde G}$ is isomorphic to $\mathcal{G}$ (as graphs), and therefore $\mathcal{\tilde G}$ is an unweighted network such that  if we reorder its nodes by using the centrality of each node we recover ranking $r$, and therefore the inverse ranking problem has positive solution for every ranking (possibly with ties).
\end{proof}

Finally, we will consider the following strict version of the Inverse ranking problem that only consider rankings without ties, i.e. permutations of the nodes.

\begin{problem}[Inverse strict ranking problem for unweighted networks]\label{prob:strict}
Given a centrality measure such as eigenvector or PageRank centrality and a ranking (without ties) of the nodes of the network, find an unweighted network $\mathcal{G}=(\mathcal{N},\mathcal{E})$ such that  if we reorder its nodes by using the centrality of each node (descent ordering), then we recover the prefixed  ranking.
\end{problem}

It is easy to prove a similar result to Theorem~\ref{thm:equivalence} for this case that ensures that there is a positive answer to the Inverse strict ranking problem for unweighted networks if and only if we can find an unweighted network whose centrality mesure corresponds to the $(1,1,\cdots,1)$-ranking.

\section{The inverse ranking problem for undirected networks}\label{sec:und-loop}

In this section we will look for an unweighed and undirected network such that its (eigenvector or PageRank) centrality produce any prescribed ranking of nodes. We will consider two different cases in terms of the existence of loops or not in the network.  Note that if loops are not allowed in the undirected graphs then the Inverse ranking problem has no solution in general, as the following example shows.

\begin{example}
If we consider the case of unweighted and undirected networks of $n=4$ nodes and without ties, in this case  there are only 6 types of connected graphs with no loops up to isomorphism, as Figure~\ref{fig:Ejemplo4} shows. If we compute the node $(r_1,,\cdots,r_k)$-rankings for the eigenvector centrality $\mathbf{c}=(c_1,c_2,c_3,c_4)^t$, we only get the following types of rankings:
\[
c_1>c_2=c_3=c_4,\\
c_1=c_2>c_3=c_4,\\
c_1>c_2=c_3>c_4\,\
c_1=c_2=c_3=c_4,
\]
but the number of $(r_1,,\cdots,r_k)$-rankings in this case is $2^3=8$, so there are some rankings that cannot be obtained for the eigenvector centrality of a unweighted and undirected networks of $n=4$ nodes and without ties.

\begin{figure}
\begin{minipage}[c]{0.32\textwidth} 
\begin{center}
\begin{tikzpicture}[node distance={15mm}, thick, main/.style = {draw, circle}]
...main/.style = {draw, circle}] 
\node[main] (1) {$1$};
\node[main] (2) [right of=1]{$2$};
\node[main] (3) [below of=1]{$3$};
\node[main] (4) [right of=3]{$4$};
\draw (1) -- (2);
\draw (1) -- (3);
\draw (1) -- (4); 
\end{tikzpicture}
\[
c_1>c_2=c_3=c_4
\]
\end{center} 
\end{minipage}
\hfill
\begin{minipage}[c]{0.32\textwidth} 
\begin{center}
\begin{tikzpicture}[node distance={15mm}, thick, main/.style = {draw, circle}]
...main/.style = {draw, circle}] 
\node[main] (1) {$1$};
\node[main] (2) [right of=1]{$2$};
\node[main] (3) [below of=1]{$3$};
\node[main] (4) [right of=3]{$4$};
\draw (1) -- (2);
\draw (1) -- (3);
\draw (2) -- (4); 
\end{tikzpicture}
\[
c_1=c_2>c_3=c_4
\]
\end{center} 
\end{minipage}
\hfill
\begin{minipage}[c]{0.32\textwidth} 
\begin{center}
\begin{tikzpicture}[node distance={15mm}, thick, main/.style = {draw, circle}]
...main/.style = {draw, circle}] 
\node[main] (1) {$1$};
\node[main] (2) [right of=1]{$2$};
\node[main] (3) [below of=1]{$3$};
\node[main] (4) [right of=3]{$4$};
\draw (1) -- (2);
\draw (1) -- (3);
\draw (2) -- (4);
\draw (3) -- (4);  
\end{tikzpicture}
\[
c_1=c_2=c_3=c_4
\]
\end{center} 
\end{minipage}
\[\,\]
\begin{minipage}[c]{0.32\textwidth} 
\begin{center}
\begin{tikzpicture}[node distance={15mm}, thick, main/.style = {draw, circle}]
...main/.style = {draw, circle}] 
\node[main] (1) {$1$};
\node[main] (2) [right of=1]{$2$};
\node[main] (3) [below of=1]{$3$};
\node[main] (4) [right of=3]{$4$};
\draw (1) -- (2);
\draw (1) -- (3);
\draw (1) -- (4);
\draw (2) -- (3);  
\end{tikzpicture}
\[
c_1>c_2=c_3>c_4
\]
\end{center} 
\end{minipage}
\hfill
\begin{minipage}[c]{0.32\textwidth} 
\begin{center}
\begin{tikzpicture}[node distance={15mm}, thick, main/.style = {draw, circle}]
...main/.style = {draw, circle}] 
\node[main] (1) {$1$};
\node[main] (2) [right of=1]{$2$};
\node[main] (3) [below of=1]{$3$};
\node[main] (4) [right of=3]{$4$};
\draw (1) -- (2);
\draw (1) -- (3);
\draw (1) -- (4);
\draw (2) -- (3);
\draw (2) -- (4);  
\end{tikzpicture}
\[
c_1=c_2>c_3=c_4
\]
\end{center} 
\end{minipage}
\hfill
\begin{minipage}[c]{0.32\textwidth} 
\begin{center}
\begin{tikzpicture}[node distance={15mm}, thick, main/.style = {draw, circle}]
...main/.style = {draw, circle}] 
\node[main] (1) {$1$};
\node[main] (2) [right of=1]{$2$};
\node[main] (3) [below of=1]{$3$};
\node[main] (4) [right of=3]{$4$};
\draw (1) -- (2);
\draw (1) -- (3);
\draw (1) -- (4);
\draw (2) -- (3);
\draw (2) -- (4);
\draw (3) -- (4);  
\end{tikzpicture}
\[
c_1=c_2=c_3=c_4
\]
\end{center} 
\end{minipage}
\caption{All possible rankings (eventually with ties) for the eigenvector centrality of a connected, undirected and without loops graphs of $4$ nodes.}
\label{fig:Ejemplo4}
\end{figure}
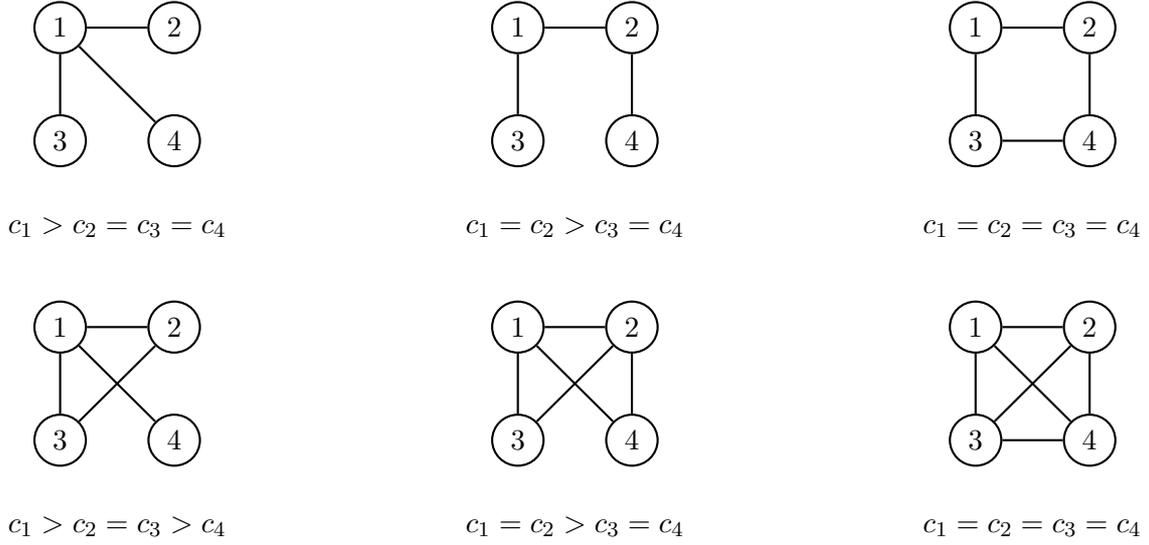
\end{example}

Despite the fact that there is not a positive answer to the inverse ranking problem for undirected and without ties (as the previous example shows) , the situation changes if we allow loops in the networks. We will prove this fact by constructing explicitly a family of network unweighted and undirected network with ties whose eigenvector or (classic) PageRank centrality produces any node ranking that we wish. In order to describe such family of networks, we first need to introduce some notation. 

For every $k,m\in\N$ we consider the matrices
\begin{equation}
\label{eq:notation}
\mathbf{1}_{k,m}=\left(
                     \begin{array}{ccc}
                       1 & \dots & 1 \\
                       \vdots & \ddots & \vdots \\
                       1 & \dots & 1 \\
                     \end{array}
                   \right)\in M_{n\times m}(\R),\qquad \mathbf{0}_{k,m}=\left(
                     \begin{array}{ccc}
                       0 & \dots & 0 \\
                       \vdots & \ddots & \vdots \\
                       0 & \dots & 0 \\
                     \end{array}
                   \right)\in M_{n\times m}(\R).
\end{equation}
If $k=m$, then we will denote $\mathbf{1}_{k,k}=\mathbf{1}_{k}$ and $\mathbf{0}_{k,k}=\mathbf{0}_{k}$. By using this notation we construct the following networks:

\begin{definition}\label{und-loop-matrix}
Given $n\in\N$ and  a list of integers $(r_1,\cdots r_k)$ such that $1\le r_1,\cdots,r_k\le n$  and $r_1+\cdots+r_k=n$, we consider the unweighted, undirected network of $n$ nodes and with ties ${\mathcal G}(r_1,\cdots, r_k)=({\mathcal N}, {\mathcal E})$ whose adjacency matrix is defined by blocks as
\[
A(r_1,\cdots, r_k)=A=\left(
        \begin{array}{ccccc}
            \mathbf{1}_{r_1}              &\mathbf{1}_{r_1,r_2}     &\dots    & \mathbf{1}_{r_1,r_{k-1}}    & \mathbf{1}_{r_1,r_k}  \\
            \cline{5-1} \mathbf{1}_{r_2,r_1}    & \mathbf{1}_{r_2}    &\dots    &\multicolumn{1}{c|}{ \mathbf{1}_{r_2,r_{k-1}}}  &\mathbf{0} _{r_2,r_k}  \\
            \cline{4-2}\vdots  &\hss  & \multicolumn{1}{c|}{ \iddots} &\iddots    &\vdots\\
            \cline{3-3}\vdots  & \multicolumn{1}{c|}{ \mathbf{1}_{r_{k-1},r_2}}  &   \iddots& \mathbf{0}_{r_{k-1}}    &\mathbf{0}_{r_{k-1},r_k}  \\
            \cline{2-1} \multicolumn{1}{c|}{ \mathbf{1}_{r_k,r_1}} &\mathbf{0}_{r_k,r_2}   &\dots     &   \mathbf{0}_{r_{k-1},r_k}&\mathbf{0}_{r_k}
        \end{array}
 \right)\in M_{n\times n}(\R).
\]
\end{definition}

It is clear that for every list of integers $(r_1,\cdots r_k)$ such $1\le r_1,\cdots,r_k\le n$  such that $r_1+\cdots+r_k=n$, network ${\mathcal G}(r_1,\cdots, r_k)$ is undirected and with loops, since $A(r_1,\cdots, r_k)$ is a symmetric matrix and some of its diagonal elements are nonzero. Furthermore, the family of networks of the form ${\mathcal G}(r_1,\cdots, r_k)$ gives a positive answer to the inverse ranking problem for undirected and unweighted networks with ties, either for eigenvector and (classic) PageRank centralities, as the following Theorem shows.

\begin{theorem}\label{und-loop-thm}
Given $q\in(0,1)$, for every $n\in\N$ and  any list of integers $(r_1,\cdots r_k)$ such than $1\le r_1,\cdots,r_k\le n$  and $r_1+\cdots+r_k=n$, the network ${\mathcal G}(r_1,\cdots, r_k)$ ranks its nodes  according to its eigenvector centrality and PageRank centrality with dumping factor $q$ as the $(r_1,\cdots,r_k)$-ranking. Therefore, the inverse ranking problem has always positive solution for unweighted and undirected networks with ties.
\end{theorem}

\begin{proof}
By using Theorem~\ref{thm:equivalence}, it is enough to prove that ${\mathcal G}(r_1,\cdots, r_k)$ ranks its nodes according to the $(r_1,\cdots,r_k)$-ranking for every list of integers $(r_1,\cdots r_k)$ such than $1\le r_1,\cdots,r_k\le n$  and $r_1+\cdots+r_k=n$. Let us fix a list of integer $(r_1,\cdots r_k)$ in previous conditions. We are going to prove that if we rank the nodes of ${\mathcal{G}}(r_1,\cdots, r_k)$ according to its eigenvector or (classic) PageRank centrality, then we recover the $(r_1,\cdots,r_k)$-ranking. We consider two cases:

\medskip
\noindent$\bullet$\underbar{Ranking for  eigenvector centrality}. Notice that if  $A=A(r_1,\cdots,r_k)$ is the adjacency matrix of ${\mathcal{G}}(r_1,\cdots, r_k)$, then $A$ is symmetric since ${\mathcal{G}}(r_1,\cdots, r_k)$ is undirected. Furthermore, since ${\mathcal{G}}(r_1,\cdots, r_k)$  is strongly connected by using the classic Perron-Frobenius Theorem\cite{Meyer}, then its eigenvector centrality $\mathbf{c}\in \mathbb{R}^n$ is the unique vector such that $\mathbf{c}>0$, $\mathbf{c}^t\mathbf{e}=1$ and
\begin{equation}\label{eq:simetrica}
A\mathbf{c}=\rho(A)\mathbf{c},
\end{equation}
where $\rho(A)$ is the spectral radius of $A$. If $\mathbf{c}\in \mathbb{R}^n$ is the eigenvector centrality of ${\mathcal{G}}(r_1,\cdots, r_k)$, then we denote it as
\[
\mathbf{c}^t=(c_1^{(1)}, c_1^{(2)},\dots,c_1^{(r_1)},c_2^{(1)},\dots,c_2^{(r_2)},\dots,c_k^{(1)},\dots,c_k^{(r_k)}).
\]
In order to show that the ranking the nodes of ${\mathcal{G}}(r_1,\cdots, r_k)$ according to its eigenvector or (classic) PageRank centrality,coincides with the $(r_1,\cdots,r_k)$-ranking, the following facts will be proven:

\begin{itemize}
\item[{\it (a)}] {\em Fact}: $c_i^{(1)}=c_i^{(2)}=\dots=c_i^{(r_i)}$ for every $i=1,\dots, k$. Note that the $r_i$-block of~\nref{eq:simetrica} is
\[\left\{
    \begin{array}{ll}
      (c_1^{(1)}+\dots+c_1^{(r_1)})+\dots+(c_{k+1-i}^{(1)}+\dots+c_{k+1-i}^{(r_{k+1-i})})=&\rho(A)c_i^{(1)}\\
      (c_1^{(1)}+\dots+c_1^{(r_1)})+\dots+(c_{k+1-i}^{(1)}+\dots+c_{k+1-i}^{(r_{k+1-i})})=&\rho(A)c_i^{(2)}\\
      \vdots &\\
      (c_1^{(1)}+\dots+c_1^{(r_1)})+\dots+(c_{k+1-i}^{(1)}+\dots+c_{k+1-i}^{(r_{k+1-i})})=&\rho(A)c_i^{(r_i)}.
    \end{array}
  \right.
\]
Since the left-hand sides of these equations are equal, then we obtain that $c_i^{(1)}=c_i^{(2)}=\dots=c_i^{(r_i)}$, for all $i=1,\dots, k$.

\item[{\it (b)}] {\em Fact}: $c_1^{(1)}>c_2^{(1)}>\dots>c_1^{(r_k)}$. Now, the first equation of each block of~\nref{eq:simetrica}, give
\[
\left\{
  \begin{array}{ll}
    r_1c_1^{(1)}+\dots+r_kc_k^{(1)}=\rho(A)c_1^{(1)} \\
    r_1c_1^{(1)}+\dots+r_{k-1}c_{k-1}^{(1)}+0=\rho(A)c_2^{(1)} \\
    r_1c_1^{(1)}+\dots+r_{k-2}c_{k-2}^{(1)}+0=\rho(A)c_3^{(1)} \\
    \vdots
  \end{array}
\right.
\]
Finally, the first equation of last system minus the second one gives
\[
0<r_kc_k^{(1)}=\rho(A)(c_1^{(1)}-c_2^{(1)})
\]
so $c_1^{(1)}>c_2^{(1)}$ since $\rho(A)>0$ because $A$ is a non-null symmetric matrix\cite{HoJo}. Similarly, $c_2^{(1)}>c_3^{(1)}>\dots>c_k^{(1)}$.
\end{itemize}

\noindent$\bullet$\underbar{Ranking for the classic PageRank centrality}. If $\mathbf{c}\in \mathbb{R}^n$ is the PageRank centrality of ${\mathcal{G}}(r_1,\cdots, r_k)$ with dumpling factor $q$, we denote as before
\[
\mathbf{c}^t=(c_1^{(1)}, c_1^{(2)},\dots,c_1^{(r_1)},c_2^{(1)},\dots,c_2^{(r_2)},\dots,c_k^{(1)},\dots,c_k^{(r_k)}).
\] 
Again, in order to complete the proof, the following facts will be proven:

The theorem is proved if we show:
\begin{itemize}
\item[{\it (c)}] {\em Fact}: $c_i^{(1)}=c_i^{(2)}=\dots=c_i^{(r_i)}$ for every $i=1,\dots, k$.  If we explicitly write the equations corresponding to the $r_i$-block of~\nref{eq:PR}, i.e., the $(r_1+\dots-r_{i-1}+1)^{\rm th}$ to the $(r_1+\dots+r_i)^{\rm th}$-equations, we have
\[
\left\{
    \begin{array}{ll}
      q\left(\frac{c_1^{(1)}+\dots+c_1^{(r_1)}}{\sum_{j=1}^k r_j}+\frac{c_2^{(1)}+\dots+c_2^{(r_2)}}{\sum_{j=1}^{k-1} r_j}+\dots+\frac{c_{k+1-i}^{(1)}+\dots+c_{k+1-i}^{(r_{k+1-i})}}{\sum_{j=1}^ir_j} \right)=c_i^{(1)}-\frac{1-q}n \\
       q\left(\frac{c_1^{(1)}+\dots+c_1^{(r_1)}}{\sum_{j=1}^k r_j}+\frac{c_2^{(1)}+\dots+c_2^{(r_2)}}{\sum_{j=1}^{k-1} r_j}+\dots+\frac{c_{k+1-i}^{(1)}+\dots+c_{k+1-i}^{(r_{k+1-i})}}{\sum_{j=1}^ir_j} \right)=c_i^{(2)}-\frac{1-q}n\\
       \vdots\\
      q\left(\frac{c_1^{(1)}+\dots+c_1^{(r_1)}}{\sum_{j=1}^k r_j}+\frac{c_2^{(1)}+\dots+c_2^{(r_2)}}{\sum_{j=1}^{k-1} r_j}+\dots+\frac{c_{k+1-i}^{(1)}+\dots+c_{k+1-i}^{(r_{k+1-i})}}{\sum_{j=1}^ir_j} \right)=c_i^{(r_i)}-\frac{1-q}n.
    \end{array}
  \right.
\]
Since the left-hand side of all these equations is the always the same, we get  that $c_i^{(1)}=c_i^{(2)}=\dots=c_i^{(r_i)}$, $i=1,\dots,k$.

\item[{\it (d)}] {\em Fact}: $c_1^{(1)}>c_2^{(1)}>\dots>c_1^{(r_k)}$. Now if we write the first equation of each block of~\nref{eq:PR}, i.e, the first, the $(r_1+1)^{\rm th}$, the $(r_1+r_2+1)^{\rm th}$-equation... and use previous linear system, we obtain
\[
\left\{
  \begin{array}{ll}
    q\left(\frac{r_1c_1^{(1)}}{\sum_{j=1}^kr_j}+\dots+ \frac{r_kc_k^{(1)}}{\sum_{j=1}^1r_j}\right)=c_1^{(1)}-\frac{1-q}n\\
    q\left(\frac{r_1c_1^{(1)}}{\sum_{j=1}^kr_j}+\dots+ \frac{r_{k-1}c_{k-1}^{(1)}}{\sum_{j=1}^2r_j}+0\right)=c_2^{(1)}-\frac{1-q}n \\
    q\left(\frac{r_1c_1^{(1)}}{\sum_{j=1}^kr_j}+\dots+ \frac{r_{k-2}c_{k-2}^{(1)}}{\sum_{j=1}^3r_j}+0\right)=c_3^{(1)}-\frac{1-q}n \\
    \vdots
  \end{array}
\right.
\]
The first equation of this system minus the second one gives
\[
0<q\frac{r_kc_k^{(1)}}{\sum_{j=1}^1r_j}=c_1^{(1)}-c_2^{(1)}, \hbox{ so } c_1^{(1)}>c_2^{(1)}.
\]
Similarly, the second equation minus the third one leads to $c_2^{(1)}>c_3^{(1)}$, and so on, until we conclude that $c_1^{(1)}>c_2^{(1)}>\dots>c_1^{(r_k)}$.
\end{itemize}
\end{proof}

Let us give a numerical example that illustrate how the family of networks of the form ${\mathcal G}(r_1,\cdots, r_k)$ solves positively the Inverse ranking problem for unweighted and undirected networks with loops.

\begin{example}
If we want to construct an unweighted and undirected network with $n=6$ nodes such that its eigenvector centrality and (classic) PageRank centrality are given by the $(1,3,2)$ ranking, i.e. 
\[
c_1>c_2=c_3=c_4>c_5=c_6,
\]
then we consider the network $\mathcal{G}(1,3,2)$ whose adjacency matrix can be seen in Figure~\ref{fig:EjemploNoDirigido}.

\begin{figure}
\begin{minipage}[c]{0.49\textwidth}
\begin{center}
\begin{tikzpicture}[node distance={15mm}, thick, main/.style = {draw, circle}]
...main/.style = {draw, circle}] 
\node[main] (1) {$1$};
\node[main] (2) [right of=1]{$2$};
\node[main] (3) [below right of=2]{$3$};
\node[main] (4) [below left of=3]{$4$};
\node[main] (6) [below left of=1]{$6$};
\node[main] (5) [below right of=6]{$5$};

\draw (1) -- (2);
\draw (1) -- (3);
\draw (1) -- (4);
\draw (1) -- (5);
\draw (1) -- (6);
\draw (2) -- (3);
\draw (2) -- (4);
\draw (3) -- (4);

\draw (1) to [out=90,in=180,looseness=8] (1);
\draw (2) to [out=0,in=90,looseness=8] (2); 
\draw (3) to [out=45,in=315,looseness=8] (3);
\draw (4) to [out=0,in=270,looseness=8] (4);
\end{tikzpicture}
\end{center}
\end{minipage}
\hfill
\begin{minipage}[c]{0.49\textwidth}
\[
A(1,3,2)=\left( 
\begin{array}{c|ccc|cc}
1 & 1 & 1 & 1 & 1 & 1 \\ \hline
1 & 1 & 1 & 1 & 0 & 0 \\
1 & 1 & 1 & 1 & 0 & 0 \\
1 & 1 & 1 & 1 & 0 & 0 \\ \hline
1 & 0 & 0 & 0 & 0 & 0 \\ 
1 & 0 & 0 & 0 & 0 & 0 
\end{array}
\right)
\]
\end{minipage}
\caption{ The undirected  graph (with loops) of six nodes $\mathcal{G}(1,3,2)$, i.e. $\mathcal{G}(r_1,r_2,r_3)$ with $r_1=1$, $r_2=3$ and $r_3=2$.}
 \label{fig:EjemploNoDirigido}
\end{figure}  
For this network $\mathcal{G}(1,3,2)$ its eigenvector centrality is
\[
c_1= 0.24197646, \qquad
c_2=c_3=c_4=  0.21363934,\qquad
c_5=c_6=0.05855276,
\]
while its PageRank (with damping factor $\alpha=0.85$) is
\[
c_1= 0.30259816, \qquad
c_2=c_3=c_4=  0.18722203,\qquad
c_5=c_6=0.06786788.
\]
Therefore, if we rank the nodes of ${\mathcal{G}}(1,3,2)$ according to its eigenvector or (classic) PageRank centrality with damping factor $q=0.85$, then we recover the $(1,3,2)$-ranking. 
\end{example}

\begin{remark}\label{remark}
Note that we can not expect a similar result to Theorem~\ref{und-loop-thm} for personalized pageRank with personalization vector $\mathbf{v}\ne \mathbf{e}$ that holds for every $q\in (0,1)$, since if $q\approx 0$ and $\mathbf{v}\ne \mathbf{e}$, then for every network $\mathcal{G}$ its personalized PageRank with damping factor $q$ is $\mathbf{c}\approx \mathbf{v}$, so we cannot recover all the rankings, since $\mathbf{v}\ne \mathbf{e}$ makes that some nodes has a strict difference in their PageRank centrality and therefore they must be in a fixed order in the obtained ranking.
\end{remark}

\section{The inverse ranking problem for directed networks without loops}\label{sec:direct}

In this final section we will analyze the inverse ranking problem for unweighed  but directed networks. By using the results obtained in Section~\ref{sec:und-loop}, we get directly that  the inverse ranking problem has always positive solution for unweighted and directed networks with ties, since every undirected network can be understood as a directed network with bidirectional links and therefore Theorem~\ref{und-loop-thm} gives a positive solution to the inverse ranking problem in the unweighted and directed settings, but only if loops are allowed. If we want to study the same problem but for complex networks without loops, we can consider the following family of networks:

\begin{definition}\label{direct-matrix}
Given $n\in\N$ and  a list of integers $(r_1,\cdots r_k)$ such that $1\le r_1,\cdots,r_k\le n$  and $r_1+\cdots+r_k=n$, we consider the unweighted, directed network of $n$ nodes and without loops of $n$ nodes $\vec{\mathcal G}(r_1,\cdots, r_k)=({\mathcal N}, {\mathcal E})$ whose adjacency matrix is defined by blocks as
\[
\vec{A}(r_1,\cdots,r_k)=A=\left(
        \begin{array}{ccccc}
            \mathbf{1}_{r_1}-I_{r_1}              &\multicolumn{1}{|c}{\mathbf{1}_{r_1,r_2}}       & \mathbf{1}_{r_1,r_3} &\dots    & \mathbf{1} _{r_1,r_k} \\
            \cline{1-5} \multicolumn{1}{c|}{\mathbf{1}}    & \mathbf{1}_{r_2}-I_{r_2}    & \mathbf{0}_{r_2,r_3}   &\dots  &\mathbf{0}_{r_2,r_k}  \\
            \cline{2-2} {\mathbf{1}_{r_3,r_1}} &\multicolumn{1}{c|}{\mathbf{1}_{r_3,r_2}}   & \mathbf{1}_{r_3}-I_{r_3}  &\ddots    &\vdots\\
            \vdots  & \hss  &  \ddots& \mathbf{1}_{r_{k-1}}-I_{r_{k-1}}   &\mathbf{0}_{r_{k-1},r_k}  \\
            \cline{4-4} \mathbf{1}_{r_k,r_1} &\mathbf{1}_{r_k,r_2}   &\dots     &   \mathbf{1}_{r_k,r_{k-1}}&\multicolumn{1}{|c}{\mathbf{1}_{r_k}-I_{r_k}}
        \end{array}
 \right)\in M_{n\times n}(\R),
\]
where $\mathbf{1}_{k,m}$, $\mathbf{0}_{k,m}$, $\mathbf{1}_{k}$ and $\mathbf{0}_{k}$ where defined in~\nref{eq:notation} and $I_k$ is the $k\times k$ identity matrix.
\end{definition}

Note that for every list of integers $(r_1,\cdots r_k)$ such $1\le r_1,\cdots,r_k\le n$  such that $r_1+\cdots+r_k=n$, network $\vec{\mathcal G}(r_1,\cdots, r_k)$ has no loops, since $\vec{A}(r_1,\cdots, r_k)$ has null diagonal. In addition to this, the family of networks of the form $\vec{\mathcal G}(r_1,\cdots, r_k)$ gives a positive answer to the inverse ranking problem for directed networks without loops for (classic) PageRank centralities, as the following Theorem shows.

\begin{theorem}\label{thm:dirigido}
Given $q\in(0,1)$, for every $n\in\N$ and  any list of integers $(r_1,\cdots r_k)$ such than $1\le r_1,\cdots,r_k\le n$, $k\ge 3$  and $r_1+\cdots+r_k=n$, network $\vec{\mathcal G}(r_1,\cdots, r_k)$ ranks its nodes  according to its (classic) PageRank centrality with dumping factor $q$ as the $(r_1,\cdots,r_k)$-ranking. Therefore, the inverse ranking problem has always positive solution for unweighted and undirected networks without loops for (classic) pageRank centrality.
\end{theorem}

\begin{proof} 
If $\mathbf{c}\in \mathbb{R}^n$ is the PageRank centrality of $\vec{\mathcal{G}}(r_1,\cdots, r_k)$ with dumping factor $q$, let us denote
\[
\mathbf{c}^t=(c_1^{(1)}, c_1^{(2)},\dots,c_1^{(r_1)},c_2^{(1)},\dots,c_2^{(r_2)},\dots,c_k^{(1)},\dots,c_k^{(r_k)}).
\] 
If we plug the block structure given by the adjacency matrix of $\vec{\mathcal{G}}(r_1,\cdots, r_k)$ in expresion~\nref{eq:PR}, the equations corresponding to its $r_i$-block
give a system of equations whose left-hand side is always of the form
\[
\begin{array}{ll}
    q\left(\frac{c_1^{(1)}+\dots+c_1^{(r_1)}}{\sum_{j=1}^k r_j-1}+\frac{c_i^{(1)}+\dots+c_i^{(r_i)}-c_i^{(i)}}{\sum_{j=1}^i r_j-1}+\frac{c_{i+1}^{(1)}+\dots+c_{i+1}^{(r_{i+1})}}{\sum_{j=1}^{i+1} r_j-1}+\dots+\frac{c_k^{(1)}+\dots+c_k^{(r_k)}}{\sum_{j=1}^k r_j-1}\right),
\end{array}
\]
while the right-hand sides are $c_i^{(1)}-\frac{1-q}n$, $c_i^{(2)}-\frac{1-q}n$, \dots, $c_i^{(r_i)}-\frac{1-q}n$ respectively. Therefore, since the left-hand sides are equal, we get that $c_i^{(1)}=c_i^{(2)}=\dots=c_i^{(r_i)}$.  Moreover, the first equations of each block  in expression~\nref{eq:PR} together with the equality of the $c_i$'s we have just proved give
\begin{equation}\label{eq:system}
\left\{
  \begin{array}{ll}
     q\left(\frac{(r_1-1)c_1^{(1)}}{\sum_{j=1}^kr_j-1}+\frac{r_2c_2^{(1)}}{\sum_{j=1}^2r_j-1}+\dots+\frac{r_kc_k^{(1)}}{\sum_{j=1}^kr_j-1}\right)=c_1^{(1)}-\frac{1-q}n\\
     q\left(\frac{r_1c_1^{(1)}}{\sum_{j=1}^kr_j-1}+\frac{(r_2-1)c_2^{(1)}}{\sum_{j=1}^2r_j-1}+\frac{r_3c_3^{(1)}}{\sum_{j=1}^3r_j-1}+\dots+\frac{r_kc_k^{(1)}}{\sum_{j=1}^kr_j-1}\right)=c_2^{(1)}-\frac{1-q}n\\
     q\left(\frac{r_1c_1^{(1)}}{\sum_{j=1}^kr_j-1}+\frac{(r_3-1)c_3^{(1)}}{\sum_{j=1}^3r_j-1}+\frac{r_4c_4^{(1)}}{\sum_{j=1}^4r_j-1}+\dots+\frac{r_kc_k^{(1)}}{\sum_{j=1}^kr_j-1}\right)=c_3^{(1)}-\frac{1-q}n\\
     \vdots\\
    q\left(\frac{r_1c_1^{(1)}}{\sum_{j=1}^kr_j-1}+\frac{(r_k-1)c_k^{(1)}}{\sum_{j=1}^kr_j-1}\right)=c_k^{(1)}-\frac{1-q}n
  \end{array}
\right.
\end{equation}
Now, by subtracting the second equation minus the third equation, the third equation minus the forth one, etc. in~\nref{eq:system}, we obtain that
\[
0<q\left(\frac{(r_i-1)c_i^{(1)}}{\sum_{j=1}^ir_j-1}+\frac{c_{i+1}^{(1)}}{\sum_{j=1}^{i+1}r_j-1}\right)=c_i^{(1)}-c_{i+1}^{(1)}, \qquad i=2,\dots, k-1,
\]
so $c_i^{(1)}>c_{i+1}^{(1)}$ for every $i=2,\dots, k-1$.

On the other hand, if we compute  the first equation minus the second one in~\nref{eq:system}, we get that
\[
q\left(\frac{c_2^{(1)}}{\sum_{j=1}^2r_j-1}-\frac{c_1^{(1)}}{\sum_{j=1}^kr_j-1}\right)=c_1^{(1)}-c_2^{(1)},
\]
which implies that
\[
\left({1+\frac{q}{\sum_{j=1}^kr_j-1}}\right)c_1^{(1)}=\left(1+\frac{q}{r_1+r_2-1}\right) c_2^{(1)},
\]
so $c_1^{(1)}>c_2^{(2)}$ as soon as
\[
\sum_{j=1}^kr_j>r_1+r_2,
\] 
which is clearly true under the hypothesis $k\ge 3$ and therefore we conclude that 
\[
c_1^{(1)}= c_1^{(2)}=\dots=c_1^{(r_1)}>c_2^{(1)}= c_2^{(2)}=\dots=c_2^{(r_2)}>\dots>c_k^{(1)}= c_k^{(2)}=\dots=c_k^{(r_k)}.
\]
\end{proof}

Let us give a numerical example that illustrate how the family of networks of the form $\vec{\mathcal{G}}(r_1,\cdots, r_k)$ solves positively the inverse ranking problem for unweighted and directed networks without ties.

\begin{example}
If we want to construct a directed network with $n=6$ nodes and without ties such that its (classic) PageRank centrality are given by the $(1,3,2)$ ranking, i.e. 
\[
c_1>c_2=c_3=c_4>c_5=c_6,
\]
then we consider $\vec{\mathcal{G}}(1,3,2)$, whose adjacency matrix appears in Figure~\ref{fig:ejemploDirigido}. For this network PageRank (with damping factor $\alpha=0.85$) is
\[
c_1= 0.2267373, \qquad
c_2=c_3=c_4=  0.20671376,\qquad
c_5=c_6=0.07656071.
\]
Therefore, if we rank the nodes of ${\mathcal{G}}(1,3,2)$ according to its eigenvector or (classic) PageRank centrality with damping factor $q=0.85$, then we recover the $(1,3,2)$-ranking. 
\begin{figure}
\begin{minipage}[c]{0.49\textwidth}
\begin{center}
\begin{tikzpicture}[node distance={15mm}, thick, main/.style = {draw, circle}]
...main/.style = {draw, circle}] 
\node[main] (1) {$1$};
\node[main] (2) [right of=1]{$2$};
\node[main] (3) [below right of=2]{$3$};
\node[main] (4) [below left of=3]{$4$};
\node[main] (6) [below left of=1]{$6$};
\node[main] (5) [below right of=6]{$5$};

\draw[stealth-stealth] (1) -- (2);
\draw[stealth-stealth] (1) -- (3);
\draw[stealth-stealth] (1) -- (4);
\draw[stealth-stealth] (1) -- (5);
\draw[stealth-stealth] (1) -- (6);
\draw[stealth-stealth] (2) -- (3);
\draw[stealth-stealth] (2) -- (4);
\draw[stealth-stealth] (3) -- (4); 
\draw[stealth-stealth] (5) -- (6); 

\draw[-stealth] (5) -- (2);
\draw[-stealth] (5) -- (3);
\draw[-stealth] (5) -- (4);
\draw[-stealth] (6) -- (2);
\draw[-stealth] (6) -- (3);
\draw[-stealth] (6) -- (4);

\end{tikzpicture}
\end{center}
\end{minipage}
\hfill
\begin{minipage}[c]{0.49\textwidth}
\[
\vec{A}(1,3,2)=\left( 
\begin{array}{c|ccc|cc}
0 & 1 & 1 & 1 & 1 & 1 \\ \hline
1 & 0 & 1 & 1 & 0 & 0 \\
1 & 1 & 0 & 1 & 0 & 0 \\
1 & 1 & 1 & 0 & 0 & 0 \\ \hline
1 & 1 & 1 & 1 & 0 & 1 \\ 
1 & 1 & 1 & 1 & 1 & 0 
\end{array}
\right)
\]
\end{minipage}
\caption{The directed (without loops) graph of six nodes $\vec{\mathcal{G}}(1,3,2)$, i.e. $\vec{\mathcal{G}}(r_1,r_2,r_3)$ with $r_1=1$, $r_2=3$ and $r_3=2$.}
\label{fig:ejemploDirigido}
\end{figure}
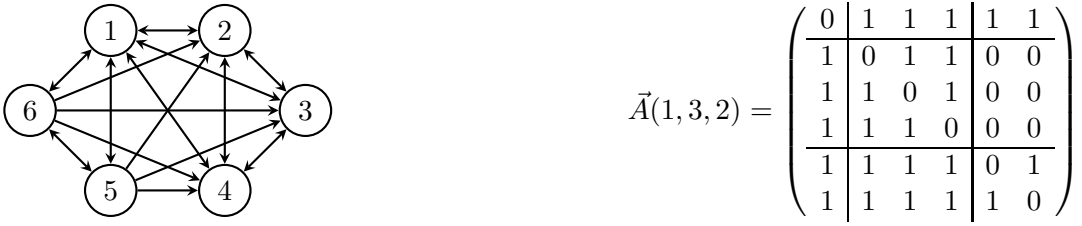

\end{example}

As it was pointed out in Remark~\ref{remark}, we can not expect a simular result to Theorem~\ref{thm:dirigido} for personalized pageRank with personalization vector $\mathbf{v}\ne \mathbf{e}$ that holds for every $q\in (0,1)$, since if $q\approx 0$ and $\mathbf{v}\ne \mathbf{e}$, then for every network $\mathcal{G}$ its personalized PageRank with damping factor $q$ is $\mathbf{c}\approx \mathbf{v}$, so we cannot recover all the rankings, since $\mathbf{v}\ne \mathbf{e}$ makes that some nodes have a strict difference in their PageRank centrality and therefore they must be in a fixed order in the obtained ranking.

\begin{remark}
Note that the inverse ranking problem is not solved in general for directed and without loops networks for eigenvector centrality by using the graph family given in Definition~\ref{direct-matrix}, since, for example, if we take $r_1=r_2=r_3=1$ we should expects that $\vec{\mathcal{G}}(1,1,1)$ satisfies that $c_1>c_2>c_3$. If we compute the eigenvector centrality of $\vec{\mathcal{G}}(1,1,1)$ by using its adjacency matrix that appears in Figure~\ref{fig:sinlazos3} then we get that
\[
c_1=0.38196594, \quad c_2=0.38196594,\quad c_3=0.23606811,
\]
so $c_1=c_2>c_3$ and we do not recover the expected $(1,1,1)$-ranking.
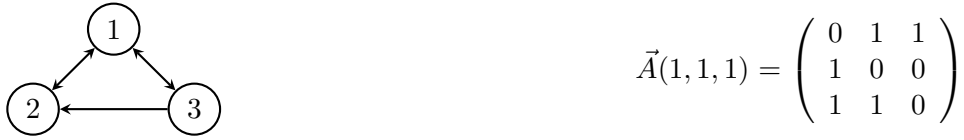
\begin{figure}
\begin{minipage}[c]{0.49\textwidth} 
\begin{center}
\begin{tikzpicture}[node distance={15mm}, thick, main/.style = {draw, circle}]
...main/.style = {draw, circle}] 
\node[main] (1) {$1$};
\node[main] (2) [below left of=1]{$2$};
\node[main] (3) [below right of=1]{$3$};
\draw[stealth-stealth] (1) -- (2);
\draw[stealth-stealth] (1) -- (3);
\draw[-stealth] (3) -- (2); 
\end{tikzpicture}
\end{center} 
\end{minipage}
\hfill
\begin{minipage}[c]{0.49\textwidth} 
\[
\vec{A}(1,1,1)=
\left(
\begin{array}{ccc}
 0 & 1 & 1\\
 1 & 0 & 0\\
 1 & 1 & 0\\
\end{array}
\right)
\]
\end{minipage}
\caption{The directed (and without loops) graph $\vec{\mathcal{G}}(1,1,1)$, i.e.  $\vec{\mathcal{G}}(r_1,r_2,r_3)$ with $r_1=r_2=r_3=1$ and its adjacency matrix.}
\label{fig:sinlazos3}
\end{figure}
\end{remark}

Although this last example shows that the graph family $\{\vec{\mathcal{G}}(r_1,\cdots,r_k)\}$ does not solve the ranking problem in general for directed and without loops networks for eigenvector centrality, we can still solve it when we only consider the {\it strict inverse ranking problem} introduced in Problem~\ref{prob:strict}, i.e., the inverse ranking problem when $r_1=r_2=\cdots= r_n=1$ and therefore the eigenvector centrality $\mathbf{c}=(c_1,c_2,\cdots,c_n)^t\in\R^n$ must fulfill that $c_1>c_2>\cdots>c_n$, as the following result shows.

\begin{theorem}
If $3\le n\in\N$, then  graph $\vec{\mathcal G}=({\mathcal N}, {\mathcal E})$ whose adjacency matrix is
\[
A=\left(
    \begin{array}{ccccc}
      0 & 1 & 0 & \dots & 0 \\
      1 & 0 & 1 & \dots & 0 \\
      \vdots & \vdots & \ddots & \ddots &   \\
      1 & 0 & \dots & 0 & 0 \\
    \end{array}
  \right)\in M_{n\times n}(\R)
\]
ranks its nodes according  to its eigenvector centrality  as the $(1,1,\cdots,1)$-ranking and the strict inverse ranking problem has always positive solution for unweighted and undirected networks without ties for eigenvector centrality.
\end{theorem}

\begin{proof} 
By using a result similar to Theorem~\ref{thm:equivalence} for the Inverse strict ranking problem, it is enough to prove that $\vec{\mathcal G}$ ranks its nodes according  to its eigenvector centrality  as the $(1,1,\cdots,1)$-ranking. 

First, note that $\vec{\mathcal G}$ is strongly connected, so by using the classic Perrron-Forbenius Theorem~\cite{Meyer}, we can compute its eigenvector centrality, that will be denoted by $\mathbf{c}=(c_1,c_2,\cdots,c_n)^t\in\R^n$. Furthermore, if we write explicitly equations given in~\nref{eq:eigen}, we get that
\[
\left\{
  \begin{array}{rl}
    c_2+c_3+\dots+c_n&=\rho(A)c_1\\
    c_1&=\rho(A)c_2\\
    c_2&=\rho(A)c_3\\
    \vdots\\
    c_{n-1}&=\rho(A)c_n
  \end{array}
\right.,
\]
where $\rho(A)$ is the spectral radius of $A$, therefore the theorem is proved from the last $(n-1)$ equations if we show that $\rho(A)>1$.  
Note that by using a result from Frobenius~\cite{Minc}, since $A$ is non-negative, then
\[
\min_i\left(\sum_{j=1}^na_{ij}\right)\le \rho(A)\le \max_i\left(\sum_{j=1}^na_{ij}\right)
\]
and if $A$ is an irreducible matrix, then equality holds on either side (and hence both sides) if and only if all row sums of $A$ are equal. If we use this result in our case, since $\vec{\mathcal G}$ is strongly connected, then $A$ is irreducible~\cite{Meyer} and the row sums of $A$ are the outdegree of each node, but
\[
gr_{out}(1)=gr_{out}(n)=1,\qquad gr_{out}(2)=\cdots=gr_{out}(n-1)=2,
\]
so by using the result of Frobenius we get that $1<\rho(A)<2$ and in particular $\rho(A)>1$, which completes the proof.
\end{proof}

Note that if we take $2\ge n\in\N$ in the last theorem, then  the Inverse strict ranking problem has no positive solution in general, for unweighted and undirected networks without loops for eigenvector centrality, simply by computing all strongly connected the unweighted and undirected networks without loops with $n=1,2$ nodes.

\section*{Acknowledgements}

This work has been partially supported by the  projects MTM2017-84194-P and  PGC2018-101625-B-I00, granted by the Spanish Ministry of Science and Innovation (AEI/FEDER, UE).



%

\end{document}